\documentclass[]{aa}
\usepackage{psfig}
\usepackage{times}
\usepackage{graphics}
\usepackage{graphicx}
\begin{document}

\title{HE0241-0155 - Evidence for a large scale homogeneous field in
a  highly magnetic white dwarf\thanks{Based on observations collected
at the European Southern Observatory, La Silla, Chile}}
\author {Dieter Reimers \inst{1}
\and Stefan Jordan \inst{2}
\and Norbert Christlieb \inst{1}}

\institute{Hamburger Sternwarte, Universit\"at Hamburg, Gojenbergsweg 112, D-21029 Hamburg, Germany
\and Institut f\"ur Astronomie und Astrophysik, Eberhard-Karls-Universit\"at T\"ubingen, Sand 1,
D-72076 T\"ubingen, Germany}
\offprints{D.~Reimers, \protect \\dreimers@hs.uni-hamburg.de}
\date{received date; accepted date}
\abstract{In the course of the Hamburg/ESO survey we have discovered a
white dwarf whose spectrum
exhibits  many similarities with the prototype of magnetic white
dwarfs Grw+70$^{\circ}$8247. In particular several  stationary line components
indicative for magnetic fields between about 150 and 400 MG are found
in both objects. However, the features between 5000 and 5500\,\AA\ in the spectrum of HE0241-0155
cannot be explained by stationary line components and demand a relatively
homogeneous magnetic field with clustering around 200\,MG. For this reason
a pure dipole model failed to reproduce this spectral region. An
offset-dipole configuration led to some improvement in the fit but a
good agreement was only possible for a geometry -- described by an
expansion into spherical harmonics -- where most of the surface is
covered with magnetic field strengths strongly clustered around
200\,MG. This may indicate the presence of a large
magnetic spot whose presence  could be tested  with time resolved
spectro-polarimetry.
}
\maketitle
\markboth{D. Reimers et al.: HE0241-0155 - Evidence for a homogeneous field}{}
\section{Introduction}
A few percent (5\%) of all known white dwarfs have detectable magnetic fields, and the
true fraction may be as high as \mbox{10 \%} if selection effects are taken into account
(Harris et al. 2003). However, the number of white dwarfs with field strengths above 100 MG,
where the spectral line pattern is no longer similar to normal H or He I spectra, is
still small, and each star is unique. Most of these stars have been detected
serendipitously during follow-up spectroscopy by surveys designed for other purposes, e.g. quasar surveys like the
Palomar-Green Survey (Schmidt \& Green 1983) or the Hamburg/ESO survey (Wisotzki
et al. 1996, Reimers \& Wisotzki, 1997). The first star of this type was
Grw+70$^{\circ}$ 8247, whose unique spectrum was first observed by
by Minkowski (1938). Its shallow spectral
features resisted a physical identification for decades.
Kemp (1970), following an idea by J. Landstreet,  proposed that a 
magnetic field would produce  circular 
dichroism, and discovered  circular
dichroism in Grw+70$^{\circ}$ 8247. Only progress in quantum
mechanical calculations of the hydrogen energy levels in strong magnetic fields made
possible an identification (Praddande, 1972, R\"osner et al. 1984). Details of the
Grw+70$^{\circ}$ 8247 spectrum and its interpretation are given by Greenstein (1984),
Angel et al. (1985) and Wickramasinghe \& Ferrario (1988). In this letter we report
on the serendipitous discovery of a near ``twin'' to Grw+70$^{\circ}$8247 in the course
of the Hamburg/ESO survey. We give a physical interpretation of its spectra
using Jordan's code (Jordan, 1988, 1992, Euchner et al. 2002) in terms of de-centered magnetic
dipole fields and we briefly discuss the remaining discrepancies between observations and theory.

\section{Observations}
HE 0251-0155 has been discovered as a quasar candidate within the Hamburg/ESO (HE) Survey
(Wisotzki et al. 1996, Reimers \& Wisotzki, 1997). The journal of spectroscopic
observations with ESO telescopes is given in Tab. 1. Wavelength and flux calibration followed
standard procedures. From the two available spectra (shown in Fig.\,ref{observation}) 
there is no evidence for variability so far.
Coordinates are RA $2^{\rm h}44^{\rm m}22\hbox{$.\!\!^{\rm s}$}0$  Dec $-01^\circ 42^\prime 41^{\prime\prime}$  (equinox 2000.0, epoche 1982.783),
 brightness is B = 16.7.

\begin{table}[h]\caption{\label{basic} \bf Journal of observations}
\rm
\begin{tabular}{llrr}
Date &          Tel./Instrument & Resolution & Exp. time  \\
\hline

13/9/98     &   1.52/B \& C     &   15 \AA\,  &  8 min \\
28/11/98    &   1.54/DFOSC      &   5 \AA\,  & 30 min \\

\end{tabular}
\end{table}

\begin{figure}[htbp]
\includegraphics[width=0.5\textwidth]{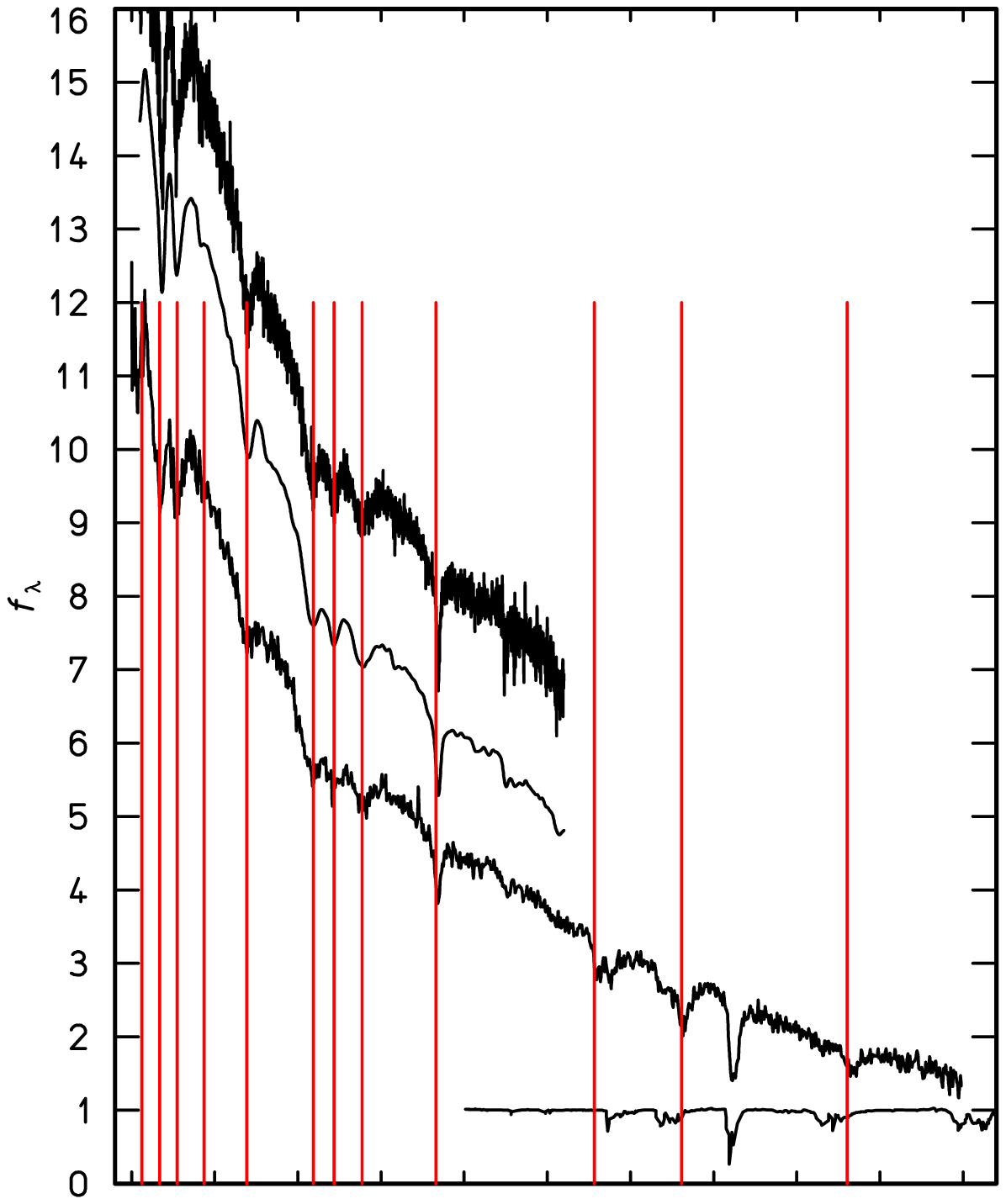}

\vspace{-1.6mm}
\includegraphics[width=0.5\textwidth]{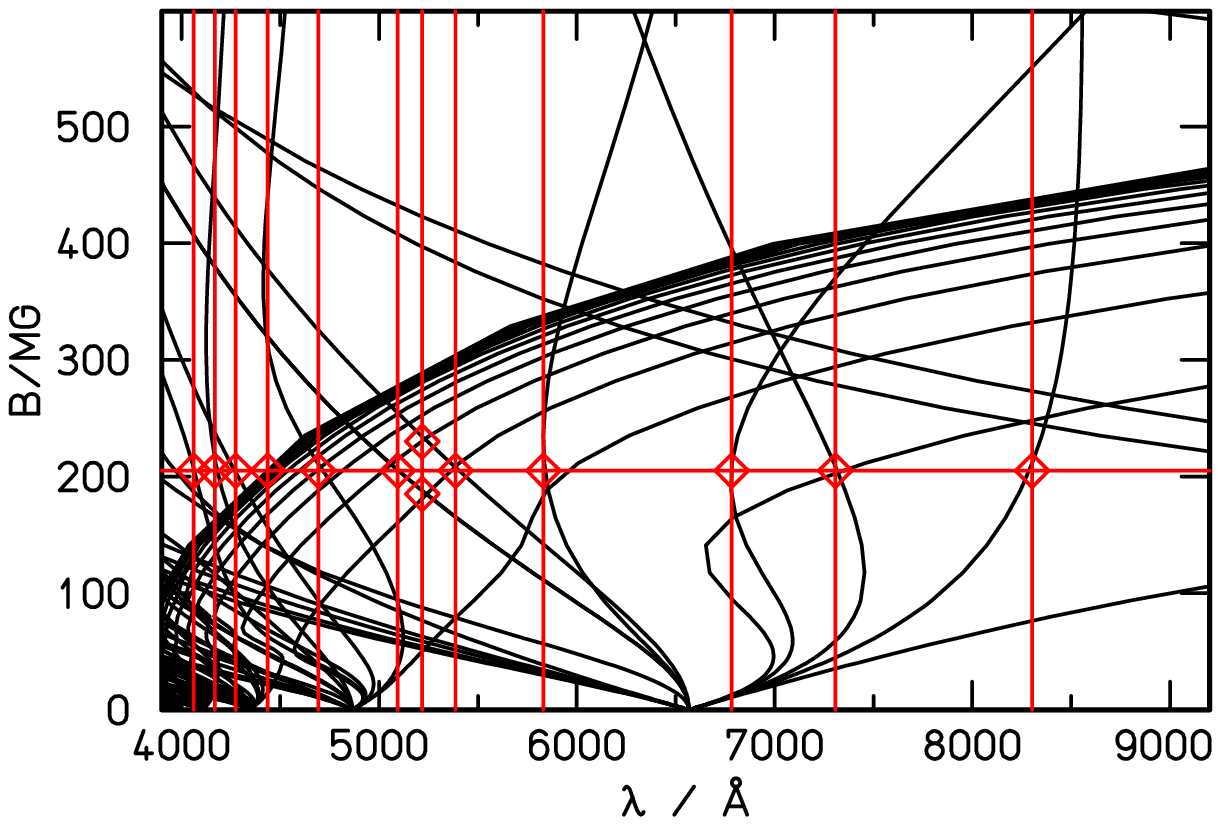}
\caption[]{Observed spectra of HE0241-0155 compared to the position of
the hydrogen line components. The upper spectrum (obtained 28/11/98) 
with a resolution of 5\,\AA\ has been 
smoothed (second curve) in order
to match the 15\,\AA\ resolution of the lower spectrum (13/9/98).
The main features arise from magnetic field
strengths of about 200\,MG (marked with circles) and are not in all cases due to stationary
line components. In order to identify the telluric features a spectrum
of HD7041 by Stevenson (1994) observed  at an airmass 1.7 is shown}
\label{observation}
\end{figure}

\begin{figure}[htbp]
\includegraphics[width=0.5\textwidth]{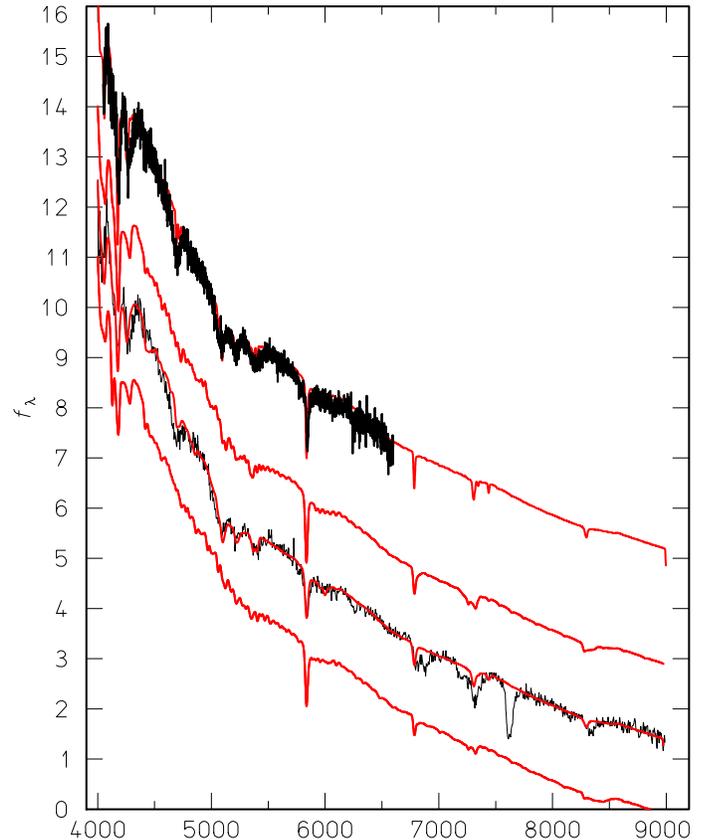}
\caption[]{Best fit for a magnetic field geometry described by
an expansion into spherical harmonics up to $l=4$ with $m=0$ 
(convolved with a 5\,\AA\ Gaussian and 
plotted over the low resolution spectrum, second
curve from below),
a pure
dipole field (lower curve), and a dipole with an offset along
the magnetic field axis (second curve from above). For comparison
the upper
curves shows the high resolution spectrum together with the
best fit model convolved with a Gaussian of 5\,\AA
}
\label{bestfit}
\end{figure}

\begin{figure}[htbp]
\includegraphics[width=0.5\textwidth]{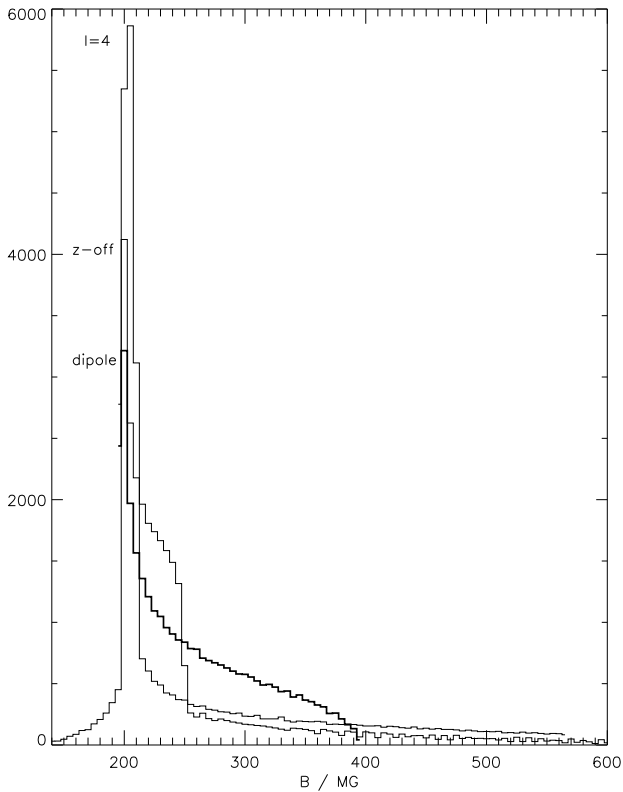}
\caption[]{Distribution of the absolute values of the magnetic field strength
over the observed hemisphere predicted by the three best fit models: the dipole model (thick line),
the off-set dipole model (medium), and the the expansion into spherical harmonics (thin)
up to $l=4$ ($m=0$)}
\label{histo}
\end{figure}

Prominent broad absorption lines are seen at 4060, 4170, 4280, 4435, 4690, 5093, 5220, 5390,
 and
5830 \AA\,. Further lines in the red are at 6780, 7310 and 8300 \AA\, (Fig.\,\ref{observation}).
Note, however, that the features in the red are partly contaminated by telluric features,
which can be identified  by the comparison spectrum of HD7041 taken by Stevenson (1994)
at high airmass, also shown in Fig.\,\ref{observation}.
A comparison with known magnetic white dwarfs shows its obvious similarity with Grw+70$^{\circ}$8247
which is a DA with a 320 MG dipole field according to Wickramasinghe \& Ferrario (1988)
and Jordan (1989).

\section{Interpretation and discussion}
The efficiency in modeling the spectra and polarization data of
magnetic white dwarfs has considerably improved due to two factors:
Firstly, instead of solving the radiative transfer equations on
thousands of surface elements on the star for each model, databases
for flux and polarization spectra computed for
a wide range of field strengths,
 viewing angles, and effective
temperatures
are used. Secondly, either evolutionary strategies 
or genetic algorithms, and a subsequent downhill
simplex algorithm (Press et al., 1992) are used 
in order to find the best fit  to the observations. The approach
used in this paper is very similar to the one described by
Euchner et al. (2002) with the exceptions that a genetic algorithm
(Caroll 2001) has been used, more viewing angles (12 instead of 9),
and an interpolation (12 $\mu$ angles) instead of a linear approximation 
are used. 

If a pure magnetic dipole is assumed, the best fit (Fig.\,\ref{bestfit})
 to the observed data was obtained
for a polar field strength of 398 MG, with the dipole axis oriented almost
($i=-86^\circ$) perpendicular to the observer, and an  effective temperature of 
12000\,K.  However, this model fails to account for the
features at 5093, 5218, and 5287\,\AA. These features do not correspond to any 
stationary line component in any reasonable range of magnetic field
 strengths; the closest agreement would have the 4f$_{-1}\rightarrow$
 2s$0$ components with a maximum at 5119\,\AA\ and 61\,MG. 
The wavelengths of stationary line components  go through maxima or minima as functions of
the magnetic field strength; they   are visible in the
spectra of magnetic white dwarfs despite a considerable variation of the field
strengths as in the case of a pure magnetic dipole (where the field strength is a factor
of two higher at the pole than at the equator, and the distribution of the magnetic field strengths
over the visible stellar surface is relatively flat (see Fig.\,\ref{histo}). 

As indicated by the circles in Fig\,\ref{observation}, the features missing in the
fit could be due to a relatively homogeneous field of about 200\,MG. One possibility
to describe a field distribution where the majority of the surface is covered by a magnetic
field close to 200\,MG is a dipole field with a large offset of the dipole relative to
the center of the star. Although the total variation of field strengths is
larger in this case, more magnetic field  strengths cluster around one relatively 
homogeneous magnetic field on one pole.

The best fit for an offset dipole (only offsets along the magnetic axis are
considered) is obtained for the same polar field strength (399 MG) and
the same orientation ($i=-86^\circ$), but for an offset of 0.16 stellar radii
away from the observer.
The histogram Fig.\,\ref{histo} shows the effect: The magnetic field strengths
are now even more strongly peaked at 200\,MG and the vast majority of surface elements has 
a magnetic field between 200 and 250\,MG. The resulting fit (upper curve of Fig.\,\ref{bestfit})
has improved considerably, and particularly the missing three line features are 
now visible -- although still somewhat too shallow -- in the theoretical model. 

Finally, a more general approach, an expansion of the magnetic field into spherical harmonics,
assuming that the source of the magnetic field lie entirely within the star
(Jacobs 1987):

\begin{displaymath}
\begin{array}{lll}
B_r=&-\sum_{l=1}^\infty \sum_{m=0}^l (l+1) (g_l^m \cos m\phi
+h_l^m \sin m \phi)\\ &\hfill  P_l^m(\cos\theta)\\
B_\theta=+ &\sum_{l=1}^\infty \sum_{m=0}^l  (g_l^m \cos m\phi
+ h_l^m \sin m \phi)\\ & \hfill dP_l^m(\cos\theta)/d\theta\\
B_\phi=&-\sum_{l=1}^\infty \sum_{m=0}^l  m (g_l^m \cos m\phi
+ h_l^m \sin m \phi)\\&\hfill dP_l^m(\cos\theta)/d\sin \theta
\end{array}
\end{displaymath}
with the associated Legendre polynomials $P_l^m$. The components
are given in spherical coordinates $r, \theta,$ and $\phi$.
For this paper we limited ourselves to $l\le 4$ and $m=0$, leading
to 4 free parameters beside the orientation of the magnetic axis 
relative to the observer. Any attempt to account for the 
20 $m \ne 0$ parameters, did only slightly improve the solution.
Since no polarization data additionally constrain the distribution of the
longitudinal component of the magnetic field, we present only the
best fit for $m=0$: $g_1^0=172$\,MG, $g_2^0=74$\,MG, $g_3^0=0.3$\,MG, $g_4^0=-0.1$\,MG, and
$i=-97^\circ$.

Fig.\,\ref{bestfit} shows  now that the overall agreement between observation
and theoretical fit is  almost perfect. The corresponding histogram is
now even stronger peaked around 200\,MG. 

It is clear that the limitation to spectra without polarization data cannot
lead to a unique solution. All distributions that cluster around 200\,MG should
lead to similar results. Such a strong clustering could be  a hint for a 
magnetic spot with a relative
homogeneous magnetic field covering a large portion of the visible hemisphere,
as discussed for the magnetic white dwarfs PG\,1031+234 (Latter et al. 1987)
 and RE\,0317-853 (Vennes et al. 2003). The origin of such a spot-like feature, is however
unclear although star spots have been found on the possible progenitors of
magnetic white dwarfs, the Ap stars, where the spots are connected to 
abundance anomalies and can be detected via Doppler imaging
(Khokhlova et al. 1986, Rice \&\ Wehlau 1994).

An extremely homogeneous field was also found in phase-resolved low-state spectra
of the polar MR Ser. Schwope et al. (1993) revealed a strongly decentered
dipolar ($z_{\rm off}\approx 0.3$ stellar radii). 

However, it must be pointed out that in Ap stars, which also have 
fossil magnetic fields, low order multipole expansions describe statistical
aspects of overall field geometry, but do not provide a realistic representation
of the actual distribution of field vectors over the surface;
the latter would be more constrained by observations of circular
polarization. Therefore,
our solution only means that the field modulus is quite uniform and large
over the visible hemisphere of the star. The relatively small region with
a higher field modulus does not strongly contribute to the strength
of the line features. 

In the framework of multipole exansions Muslimov et al. (1995) have shown that a weak quadrupole (or octupole,
etc.) component on the surface magnetic field of a white dwarf may survive the
dipole component und specific initial conditions: Particularly the evolution of
the quadrupole mode is very sensitive (via Hall effect) to the presence of
internal toroidal field. For a 0.6\,$M_{\odot}$ white dwarf with a toroidal fossil
magnetic field of strength $<10^9$\,G the dipole component declines by a factor
of three in $10^9$\,yr, while the quadrupole component is practically
unaffected. Without an internal toroidal field the dipole component still
declines by a factor of three but the quadrupole component is a factor of six
smaller after 10\,Gyr. This is much longer than the cooling age of 
a 12000\,K white dwarf, which well under 1\,Gyr for  0.6\,$M_{\odot}$ and
larger than 3\,Gyr only for masses larger than 1.2\,$M_{\odot}$.

 Whether spot-like structures could also survive
for the long cooling time of white dwarfs has to be investigated theoretically.

 If a magnetic spot is indeed present  a strong temporal variation of
the spectrum and particularly the polarization is expected, if the star
is rotating. However, no significant differences could be found between the spectra
taken in September, 1998 and November, 1998, respectively. This may be an indication
for only slow rotation as known in the case of Grw+70$^{\circ}$8247 where a rotational
period of decades or centuries is possible (Schmidt \&\ Norsworthy 1991, Friedrich \&\
Jordan 2001).
Therefore, a more detailed analysis and a decision whether a spot is present has to wait
until  time resolved spectro-polarimetry of this object becomes available.

\begin{acknowledgements}
We thank the referee John Landstreet for his very valuable comments
particularly on the situation in Ap stars. Work on magnetic white dwarfs in
T\"ubingen is supported by the DLR grant 50 OR 0201.
\end{acknowledgements}

\end{document}